\documentclass[letterpaper,11pt,english]{article}
\usepackage{graphicx}
\usepackage{hyperref}
\usepackage{amsmath}
\usepackage{authblk}
\usepackage{sidecap}

\title{Equivalence-principle Analog of the Gravitational Redshift}
\author[1]{K. Arms\thanks{kreggarms@gmail.com}}
\author[2,3]{Mario Serna\thanks{mario.serna@us.af.mil}}
\affil[1]{Department of Defense, Laurel, Maryland}
\affil[2]{United States Air Force Academy, Colorado}
\affil[3]{Air Force Research Lab, Asian Office of Aerospace Research and Development, Tokyo, Japan}

\begin{document}
	\maketitle

\begin{abstract}
What happens when two synchronized clocks on a rigid beam are both given the exact same acceleration profile?  Will they remain synchronized?  What if we use a rigid-rod Rindler acceleration profile?  The special relativity prediction surprises many people.
This experimental setup is the special-relativity analog of the gravitational redshift.
Just like two clocks higher and lower in a gravitational field lose synchronization, one sees a loss of synchronization in these clocks with `identical' acceleration profiles.
To the best of our knowledge this equivalence principle analog has never been directly measured, and current experimental techniques are sensitive enough to measure it.
We discuss the origin of the essential physics behind this synchronization loss, 
and some special conditions which simplify its experimental observation.
If validated this effect will not only test the equivalence principle from a new vantage, 
but it may one day aid in understanding and enhancing future ultra-precise navigation systems.
\end{abstract}	
	
\section{Introduction}
The equivalence principle is one of the most precisely tested phenomena in physics.  It began with Galileo's observation that all masses fall with the same acceleration.  It has survived and guided the revolutions in physics associated with special relativity,  general relativity, and quantum mechanics.   It is currently tested to $2$ parts in $10^{13}$, and experiments are planned to push this limit to $1$ part in $10^{18}$ \cite{Overduin:2012uk}. Violations are expected in many formulations of quantum gravity.
Most of the tests of the equivalence principle  study if all objects, regardless of composition, do indeed fall at the same rate.  
In this paper, we describe a different aspect of the equivalence principle that has not yet been directly tested, but as we show, can be with current to near-term technology. Finally, we observe that an appreciation for this phenomena is important for future navigation and timing.

In 1907 Einstein introduced the equivalence principle as the assumption of `the complete physical equivalence of a gravitational field and a corresponding acceleration of the reference system.' 
First we consider the physical situation in a gravitational field, then we will look at the same (similar) situation from an accelerating frame of reference.  

The gravitational redshift studies two clocks at different heights in a gravitational field.
The clock deeper in the gravitational field runs slowly compared to the clock that is higher.
Recently experiments have measured the redshift associated with less than one meter altitude difference \cite{citeulike:7894787}.
General relativity predicts for weak gravitational fields
 \begin{equation}
 \frac{\tau_h- \tau_0}{\tau_0} =  \frac{\delta f}{f_o} =  
 \frac{\Delta U}{c^2} = \frac{g\, h}{c^2}
 \label{EqGRPrediction}  
 \end{equation}
where $\tau_0$ is the time elapsed clock at a reference altitude of $0$, $\tau_h$ is the time elapsed on the clock at altitude $h$, $f_0$ and $f_h$ is the frequency of lower and upper clock respectively, $\delta f=f_h-f_0$ is the shift in the frequency observed between the clock at $h$ and at $0$, and $\Delta U \approx g \, h$ is the change in the gravitational potential. 
Observations have confirmed this expression for the gravitational redshift
to about $1$ part in $10^6$ \cite[pg 16]{lrr-2014-4}.


Although the gravitational redshift has been precisely measured, the equivalence-principle analog of the gravitational redshift does not yet appear to have been directly observed.
The equivalence-principle analog
occurs when we take two clocks on opposite ends of a `rigid' rod of size $L$ and accelerate the rod along its long axis.  We'll label the time on the left clock $\tau_L$ and the time on the right clock $\tau_R$.  
The two clocks are initially synchronized with the beam at rest ($v_0=0$).  
The rigid beam then undertakes a uniform acceleration to the right until it is traveling at a speed $v_f$.  
After acceleration, 
the equivalence principle  predicts that they will no longer be synchronized \cite{1909AnP...335....1B,romain1963time,RefOriginInertia,
PhysRevA.47.3611,2003MallinckrodAAPT}. 
In any physical test, a uniform acceleration cannot be maintained indefinitely.  
Therefore, the equivalence principle would manifest as a time difference between these two clocks before and after a period of uniform acceleration by an amount 
\begin{equation}
  \tau_R-\tau_L \approx \tau_L\,\frac{g\,L}{c^2} \approx \frac{v_f\, L}{c^2} \label{EqSRAnalogGRRedshift}
\end{equation}
where we have substituted the equivalence-principle analogs into eq(\ref{EqGRPrediction}). 
The analogs used here are $\tau_h \approx t_R$, $\tau_0 \approx \tau_L$, $\Delta U \approx g\, h \approx g L$, and 
we replaced $g\, \tau_L$ with the change in velocity $v_f-v_0=v_f$ through simple kinematics.  
We have assumed the acceleration achieved during the observation period to be non-relativistic and the time difference to be small relative to the time $\tau_L$. 
Therefore from the gravitational redshift and the equivalence principle, one expects 
two clocks that undergo the same\footnote{We will address the effect of slight differences in acceleration paths later in this paper.} acceleration to lose synchronization. 

Although the equivalence principle has been thoroughly tested \cite{lrr-2014-4}, 
we have not found a direct measurement of the 
direct equivalence principle analog of the gravitational red shift.  
The gravitational redshift tests the component of the Einstein equivalence principle known as the principle of local position invariance.   
In terms of pure physics, measuring the loss of synchronization in the above experiment would test another aspect known as the Accelerated Clock Principle (ACP) \cite{PhysRevA.47.3611} also known as the clock hypothesis.  The ACP is the assumption that 
the proper length of a space-time path equals the time evolved on a clock that follows that path.
There are several tests of the ACP in rotating systems \cite{PhysRev.129.2371}
using the K\"undig experiment.
The closest test of this phenomena that we found was in Mainwaring and Stedman\cite{PhysRevA.47.3611}.  
They mentioned the rotating tests were questionable tests of the ACP and indicated a colinear test would be ideal.  They found only one co-linear test: 
Bommel\cite{1962Bommel} had performed a linear motion test with the Mossbauer effect and piezoelectric actuators moving sources, but Baryshevsky \cite{1985Baryshevsky} cast doubt on their results.  
There are also a superficially similar few variants that generalize the Sagnac effect to detect linear motion with a closed loop of fiber \cite{2004WangPhysRevLett.93.143901}.  

We argue that measuring this direct analog is interesting for four reasons.  
First this would be a direct test or demonstration of the
classic thought experiment taught in every introduction to general relativity. 
Second, it directly tests the clock hypothesis.
Next, an experimental validation will likely yield non-obvious technological issues, and this test would be a pathfinder towards meaningful employment of next generation clocks on moving platforms.
Last, a confirmation of this effect could aid in future navigation 
technologies as the time shift along the axis of the separation of the two 
clocks will be proportional to the velocity shift along that axis.

In this paper we elaborate on this phenomena. In Sec.~\ref{SecOriginSR} we will review the derivation in special relativity.  Sec.~\ref{SecEssentialPhysics} will then decipher the essential physics of the time shift in special relativity.  
The new results begin in Sec.~\ref{SecExperimentalRealities} where
we will study if the effect survives causal cases and some experimental realities.  
Next Sec.~\ref{SecExperiments} surveys the possibilities for current technology to measure this phenomena, and 
we highlight past experiments that measure related but distinct phenomena. 

\section{The origin of the time shift in special relativity}
\label{SecOriginSR}
Rindler coordinates provide the special case where one can conveniently describe a rod of a fixed proper length (called Born rigidity) undergoing a period of constant acceleration \cite{1909AnP...335....1B,2011arXiv1105.3899F}.
\begin{SCfigure}
	\centering
	\includegraphics[width=0.65\textwidth]{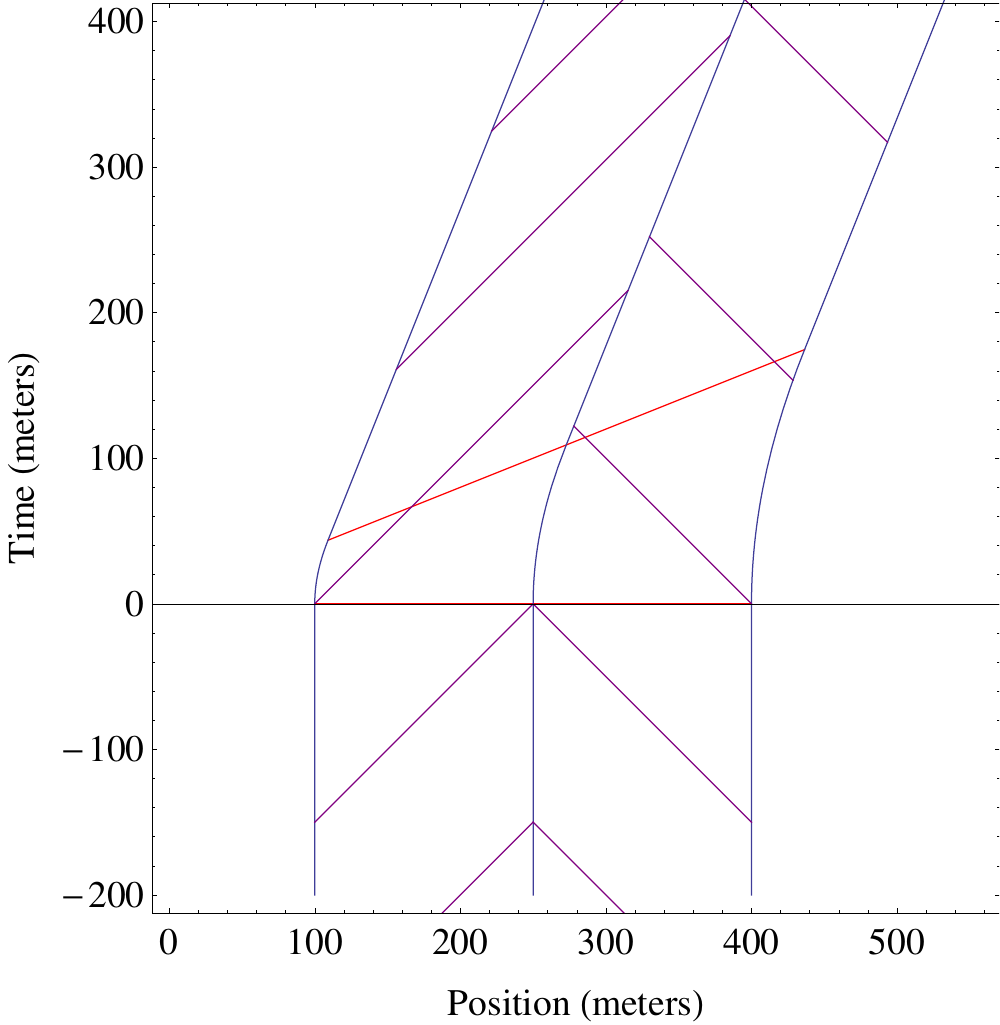}
	\caption{To see the loss of synchronization, consider a case where light leaves each side of a 300 meter rigid rod at $150$ light-meter time intervals. 
The rod undergoes acceleration that preserves Born-rigidity
from $v_0=0$ to $v_f=0.4 c$.  After acceleration, the two clocks are no longer synchronized.   }
\label{FigureTimeDifferenceDependsOnVImagePaper}
\end{SCfigure}
Figure \ref{FigureTimeDifferenceDependsOnVImagePaper} shows this loss of synchronization on a space-time diagram. 
Light pulses are being emitted toward the center of a $300$ meter rigid object at equal proper time intervals ($150$ light meters) as determined by the ACP.  
At $t<0$ one can see that the clocks $L$ and $R$ on each side of the rod are synchronized: $\tau_R-\tau_L\equiv\Delta \tau_{RL} = 0$.   At $t=0$ the rigid object begins to accelerate preserving Born rigidity.  
In figure \ref{FigureTimeDifferenceDependsOnVImagePaper} the rigid rod stops accelerating at $v_f=0.4\, c$ which is around $ t \approx 30$ light meters on the left end and $ t \approx 150$ light meters on the right end.
The two sides do not stop accelerating at the same time in the rest frame due to a shift in the surface of simultaneity as seen from the accelerating rod.   
Figure \ref{FigureTimeDifferenceDependsOnVImagePaper} also shows the surface of simultaneity when the rod's instantaneous velocity reaches $v_f=0.4\, c$, and the rod ceases to accelerate.  
Before, during, and after the acceleration the proper-length of the rigid rod remains $300$ meters \cite{2011arXiv1105.3899F}.
After the acceleration the light pulses reaching the center point indicate that the two clocks on each end are no longer synchronized.  
Calculated using Rindler coordinates, the time difference is given by
\begin{equation}
 c\, \Delta \tau_{RL} = L \tanh^{-1}\left(v_f/c\right) \approx L \,v_f/c
 \label{EqDeltaTauRindler}
\end{equation}
which agrees with eq.(\ref{EqSRAnalogGRRedshift}) for $v<<c$. 
An equivalent description makes use of the surface of simultaneity in the new frame.  
Through the exchange of light pulses, all observers moving in this frame will agree that events along that surface are simultaneous.  
Therefore, we can find the same time difference by comparing the proper time of observers calculated using the ACP along that surface of simultaneity.
We show the derivation in the appendix. This is the time shift for the equivalence principle analog of the gravitational redshift.  

For a sense of scale, a $300$ meter rod accelerating from $0$ m/sec to $10$ m/sec results in  $\Delta \tau_{RL} \approx 30 \times 10^{-15}$ sec.
Others have noted the time shift associated with spatially separated observers undergoing `similar' acceleration\cite{1909AnP...335....1B,RefOriginInertia,2003MallinckrodAAPT}.
Indeed, earlier it was shown this time shift is the same phase lag introduced by electric or magnetic fields to induce a shift in the wavefront when undergoing an acceleration \cite{2003Serna-1126-6708-2003-10-054,2005Serna-1742-6596-24-1-025}.

\section{Deciphering the essential physics}
\label{SecEssentialPhysics}
What is the key physics behind this effect?  
Many initially postulate that the time shift in the Rindler case is because the two sides undergo different accelerations.  
To maintain Born rigidity during the acceleration, the left side of the beam must accelerate slightly more than the right side.
Testing this would require extreme precision to maintain Born rigidity.  
In contrast, we show two cases of acceleration profiles that have the same first-order time shift but where both sides of the bar experience exactly the same acceleration.

\begin{SCfigure}
\includegraphics[width=0.65\textwidth]{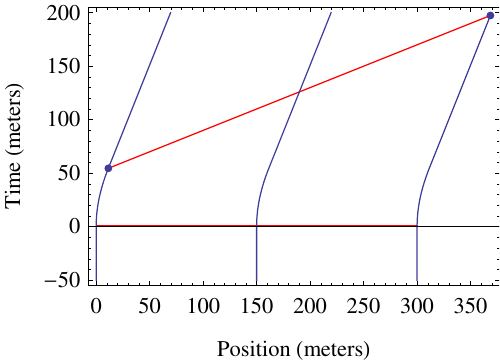}
\caption{\label{FigFiniteSameAcceleration}The two sides of the rod each experience exactly the same finite  acceleration from $v_0=0$ to $v_f$.  The proper length of the beam grows in this scenario.}
\end{SCfigure}

The first of these cases is shown in figure \ref{FigFiniteSameAcceleration}.  
Here  both sides of the beam undergo exactly the same acceleration\footnote{The authors would like to thank Neil Ashby for pointing out this special case.} 
from $v_0=0$ to $v_f=0.4 c$.  
The acceleration is completed near $t_f\approx 50$ light seconds.  
Because each point in the beam underwent the same acceleration profile, Born rigidity is not preserved, and the proper length of the beam grows to $L/\sqrt{1-(v_f/c)^2}$.  
Despite these two complications after the maneuver, the  time difference  of the clocks on opposite ends of the beam when compared to the surface of simultaneity is 
\begin{equation}
\tau_R-\tau_L = \frac{L\,v_f/c^2}{1-(v_f/c)^2 } \approx L\,v_f/c^2.
\label{EqTimeDiffSameUniformAcc} 
\end{equation}
 Every point in the bar undergoes exactly the same acceleration profile, yet the clocks still show a loss of synchronization proportional to $L\,v_f/c^2$.

\begin{SCfigure}
\includegraphics[width=0.65\textwidth]{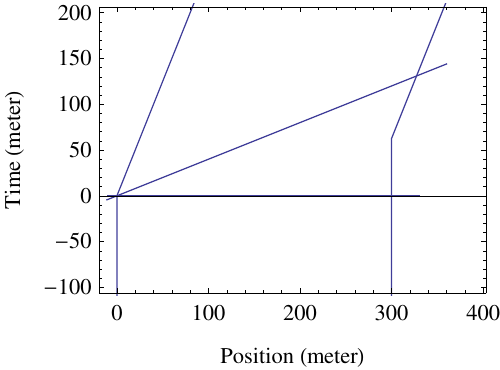}
\caption{\label{FigInstantAcceleration}The two sides of the rod each experience exactly the same instantaneous acceleration from $v_0=0$ to $v_f=0.4\,c$. Also shown is the surface of simultaneity in the final moving frame. }
\end{SCfigure}

Figure \ref{FigInstantAcceleration} shows a second case where both sides of the rod undergo the same instantaneous acceleration. 
The time chosen for the acceleration is such that the proper length of the beam is the same a few moments after the acceleration is complete as it was before the acceleration began. The resultant time shift is
\begin{equation}
\Delta \tau = \frac{2 L}{v_f} \left(1 - \sqrt{1-(v_f/c)^2} \right) \approx L\,v_f/c^2.
\end{equation}

We have shown two examples where both sides of the beam undergo exactly the same acceleration, yet the two sides shift out of synchronization by an amount $\tau_R-\tau_L \approx L\,v_f/c^2$.  This result discredits the assumption that the time shift predicted must be due to a different acceleration profile.  
For this reason we argue that the acceleration profile is not part of the essential physics.  
Instead, we argue that it lies in the shift of the surfaces of simultaneity. 
As can be seen in figures \ref{FigureTimeDifferenceDependsOnVImagePaper} to \ref{FigInstantAcceleration}, the surface of simultaneity for the initial and final frames form a wedge.  
Through almost any acceleration profile, the right side of the beam must traverse an extra proper-time interval of approximately $L\,v_f/c^2$.  
Therefore,  the loss of synchronization of the order of $L\,v_f/c^2$ is a generic feature of changes in velocity for a broad case of acceleration profiles.


\section{Experimental Considerations and Causal Cases}
\label{SecExperimentalRealities}
\label{SecExperimentRealities}

\begin{SCfigure}
	\centering	\includegraphics[width=0.35\textwidth]{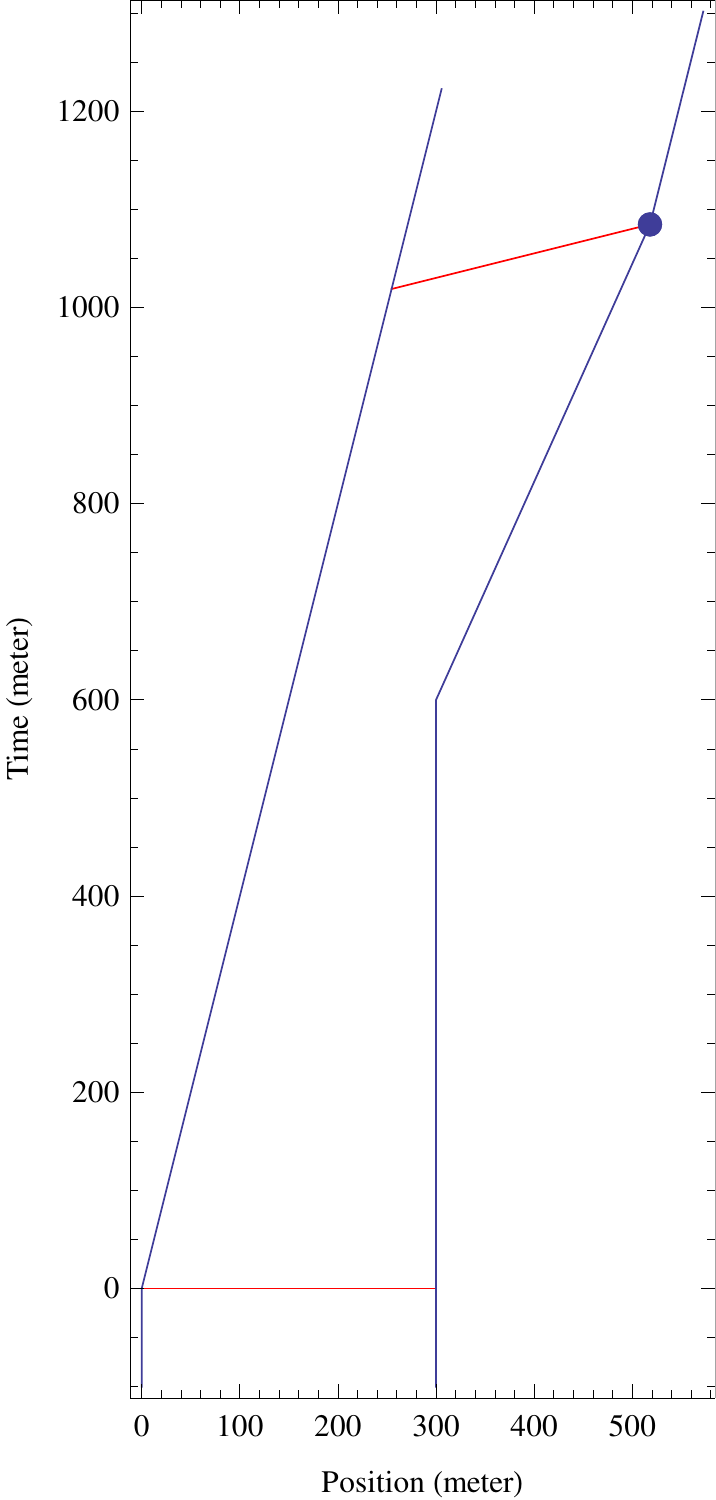}
	\caption{Space-time path for two clocks connected by a stiff rod.  The left side of the
rod receives an impulse at $t=0$ and begins to move with a velocity $v$.  The right side of the rod does not begin to move until the signal reaches it at the speed of sound $v_s$.  At the time the sound wave reaches the far end, it begins to move with a recovery speed $v_r$ until the rod reaches a new equilibrium length $L+\Delta L$.}
	\label{FigCompressedRod}
\end{SCfigure}
Although the effect is easy to calculate with perfect Born rigidity in Rindler coordinates, the setup assumes noncausal accelerations.
The system is noncausal because Born-rigidity requires the entire object to begin to accelerate non-causally at the same time $t=0$.
It also requires the entire object to non-causally stop accelerating along a surface of simultaneity.

We will now consider a more realistic scenario depicted in figure \ref{FigCompressedRod}.  This figure shows the space-time path for two clocks connected by a stiff rod.  The left side of the
rod is given a sudden push at $t=0$ and changes velocity to $v$ \cite{2011arXiv1105.3899F}.  The right side of the rod does not begin to move until the compression wave or sound signal reaches it.  
We will take the time required for the impulse on the left side to reach the right side to be $L/v_s$ where $L$ is the proper length of the rod and $v_s$ is the speed of sound \cite{2011arXiv1105.3899F}.  
At the time a sound wave reaches the far end, it begins to move with a recovery speed $v_r$ until the rod reaches a new equilibrium length $L+\Delta L$.  For non-relativistic speeds of sound and recovery velocities, this impulse approximation gives the essential behavior.
At the end of the transition when the entire rod now has proper length $L+\Delta L$ and is moving at a fixed speed $v$, the time difference between the two clocks $L$ and $R$ is now given as
\begin{eqnarray}
  \Delta \tau_{RL} & = & L\,\frac{ (\gamma_v^{-1} -1) (v_r-(\gamma_r^{-1}+1) v_s)-v (\gamma_r^{-1}+v_r v_s/c^2-1)}{v_s (v-v_r)} \nonumber \\
   & & -\Delta L \, \frac{ \left(\frac{1}{\gamma_v \gamma_r}+v\,v_r/c^2 -1 \right)}{v-v_r}
   \label{EqDeltaTauGeneral}
\end{eqnarray}
where $\gamma_v = (1 - (v/c)^2)^{-1/2}$ and $\gamma_r = (1 - (v_r/c)^2)^{-1/2}$.  The derivation is found in the appendix.
It is interesting\footnote{This is the case studied by Ref \cite{2014Natario}.  This can be partially understood because for a light-like recovery-path no proper time is accumulated during the transition from the old velocity to the new velocity.} to note that if $v_r=v_s=c$ then $\tau_{BA}=0$.   
To first order in the unitless small parameters $v/c$, $v_r/c$, $v_s/c$ we have
\begin{equation}
   \Delta \tau_{BA} \approx \frac{L \,v\,}{c^2} \left( 1 - \frac{v_r}{2 v_s } - \frac{\Delta L}{2\,L} (1 + \frac{v_r}{v}) \right) + \ldots.  \label{EqTaylorExpansionTimeDifference}
\end{equation}

The model in Figure \ref{FigCompressedRod} involves unphysical instantaneous changes in velocity.  Any real change in velocity can be modeled by the limit of many small instantaneous changes of the type considered.  Therefore, this extreme scenario captures the essential features needed to understand a realistic model.

With this in mind, equation \ref{EqTaylorExpansionTimeDifference} has some important implications for the tolerance of an experiment to observe the induced time shift.  
First if we are dealing with something like a steel cruise ship, the speed of sound in steel is $v_s = 4300$ meters/sec.  
Therefore for typical boat speeds and accelerations $v_r/v_s <<1$ and $v/v_s<<1$ and both can be neglected.  
Next once the boat changes velocity, the length of the boat will change due to the drag force of the water.  If $\frac{\Delta L}{L} << 1$ and $v_r/v \approx 1$, then the last terms are also unimportant.
In the setup described this means the time shift is dominated by $c\,\Delta \tau_{BA} \approx L\,v/c$,  a quick estimate of a steel boat suggest a $300$ meter boat may be compressed by about $6$ mm once the boat is moving at $10$ m/s 
through the water.\footnote{A typical cruise ship travels at about $20$ knots which is equal to about $10$ m/s.}  Such a $\Delta L / L$  will not alter our prediction of the time shift induced by Born rigidity.

\section{The Potential for Experimental Observation}
\label{SecExperiments}

Experimental tests of the equivalence principle are reviewed in Ref.~\cite{lrr-2014-4}, and 
the effects of linear velocity on time dilation are reviewed in Ref.~\cite{Reinhardt:2007zzb}.
In none of the past experimental tests that we could find 
were two clocks placed on ends of a rigid rod which was then linearly accelerated and then compared.
Comparing this result to gravitational time dilation would be the first direct test of the equivalence principle and the accelerated clock principle (ACP).

To measure this effect we need to have two clocks, one   
on each side of the rod with frequency width $\Delta f$ that satisfies $\Delta f / f << a L / c^2 $ where 
$a=\Delta v / \Delta \tau$ is the acceleration 
scale of the system, $L$ is the separation of the two clocks, $c$ is the speed of light, and $f$ is frequency of the clock standard.  For a separation of $L=100$ meters, an acceleration of $a=1$ meter/sec, we need a quality factor $f/\Delta f$ of more than $10^{15}$.

Experimental techniques with this scale of accuracy have recently been demonstrated \cite{citeulike:7894787,2013arXiv1305.5869H,gibney2015hyper}.
Extremely narrow line-width lasers have been made possible by the use of frequency combs \cite{diddams2000direct}.
There are even efforts to make portable narrow-line-width lasers which would also be helpful for such a demonstration \cite{Vogt2011}. 
The lasers on each end of the accelerating rod need to be independent clocks.  
The two beams can be brought together to interfere near the center.  
As the rod accelerates, the beat frequency will shift marking the time shift described above.\footnote{The authors would like to acknowledge helpful conversations with Scott Diddams, Steven Cundiff, and Ross Spencer concerning the viability of performing such an experiment.}

The closest analog of the proposed test is the validation of the transverse Doppler effect in the K$\rm{\ddot{u}}$ndig experiment \cite{PhysRev.129.2371}.    
In the K$\rm{\ddot{u}}$ndig experiment a rigid disk is rotated with a source at the center and a detector at the radius.  
The `rigid' rod essentially is only accelerating on one end.  
Recent analysis of K$\rm{\ddot{u}}$ndig experiments show some possible deviations 
from theoretical predictions which make more interesting 
this direct test of the equivalence principle\cite{1402-4896-79-6-065007}.


The proposed experiment may be confused with the generalized Sangac effect \cite{PhysRevLett.93.143901} where a segment of a fiber loop enclosing 
no area is accelerated and shows a phase shift $\Delta \phi = 4 \pi \oint_l \vec{v} \cdot d\vec{l}/c \lambda$.  
The frequency shift we describe is different. 
In the generalized Sangac effect there is only one clock which sends light signals both ways around a fiber.  
The integrated phase shift after the acceleration is complete disappears if the  entire apparatus is accelerated together rather than having a segment of the fiber undergo acceleration.  
The patent filed by Wang \emph{et.al.} \cite{wang2009stand} is nearing an example of an apparatus that may test the equivalence principle if the clocks on opposite ends of the rod are locked to a local reference standard, \emph{e.g.} Cesium or similar, such that they are comparing the proper time on the two ends of the rod. 
The patent as written appears to call for a single laser, which fails on the requirement to have two independent clocks to measure the accelerated clock hypothesis.  If the laser line width $\Delta f$  is large 
compared to the integrated shift being measured, $\Delta \phi\, c /\lambda$, then their apparatus will record no observable fringes.  If instead their linewidth is sufficiently small, then the two ends of the laser cavity cannot both be simultaneously in resonance during the acceleration.

Our purpose here is not to specify the precise experimental approach, 
but rather to simply observe that this direct equivalence principle analog 
of the gravitational redshift has never been experimentally observed to the best of our knowledge. Further, some elegant features help protect the effect from being dwarfed by experimental complications, and current technology should allow experimental observation.

\section{Conclusion}
Although time dilation has been repeatedly measured in a gravitational field, 
the equivalence-principle analog from 
linear acceleration has not yet been directly measured to the best of our knowledge. 
A general feature of such an experiment is a time shift $\Delta \tau_{RL} \approx L\,v/c^2$ in clocks at the two ends of the rod proportional 
to the change in velocity of the rod and proportional to the separation distance between the clocks. 
In section \ref{SecExperimentRealities}, we show that this time shift is  
tolerant to experimental realities.   
We have shown that this effect is a general feature for a rod whose length changes slowly compared to the speed of light ($v_r << c$).
The experimental accuracy
required to measure this relativity phenomena is just now appearing.
In a future era of ultra-precise clocks, such time differences will matter for two-way time transfers.
Because to first order the time shift is a proxy for the velocity, this effect may one day have implications for precise navigation.  
After accounting for gravitational redshifts, one could in-principle determine the velocity relative to your starting velocity by measuring the time shift of two spatially separated clocks located on extreme ends of a boat, aircraft, or other vehicle.

\section*{Acknowledgments}

We would like to thank Richard Cook, Scott Diddams, and Neil Ashby for comments on the manuscript.

\bibliographystyle{unsrt}
\bibliography{VelocityTimeShift}

\appendix
\section*{Appendix: Derivation of Key Equations}

To derive eq.~\ref{EqDeltaTauRindler} we follow two paths of constant acceleration as described in Rindler coordinates representing different ends of a rigid rod starting at $t=\tau=0$ and ending when the rigid rod has reached a velocity $v$.
Rindler coordinate are given by $ ( x(\tau), c\,t(\tau) ) = \left( c^2 a^{-1} \cosh( a \tau/c ) , c^2 a^{-1} \sinh ( a \tau/c ) \right)$. 
An object moving along an accelerating path has a 
velocity $v(\tau)/c = \tanh (a \tau/c)$.
The left-side clock which starts at position $(x_L,0)$ follows the trajectory with acceleration $a_L=c^2/x_L$.  
The clock on the right-side of the rigid rod of proper length $L$ starts at position $(x_L+L,0)$ follows the trajectory with acceleration $a_R=c^2/(x_L+L)$.
When the left side reaches a velocity $v$, the proper time on that clock is $\tau_L=x_L/c\, \tanh^{-1}(v/c)$.
When the right side reaches a velocity $v$, the proper time on that clock is $\tau_R=(x_L+L)/c\,\tanh^{-1}(v/c)$.
The proper time difference is $\tau_R-\tau_L=L/c \, \tanh^{-1}(v/c)$.


To derive eq.~\ref{EqDeltaTauGeneral} we parameterize the paths of the two clocks and then calculate the proper time elapsed between surfaces of simultaneity for the initial and final velocities.
For clock A on the left the path is given by
\begin{equation}
  (x_A(s),\ t_A(s)) =  \left\{
              \begin{array}{cc}
                 (0, \, s)& \mbox{if } s \leq 0  \\
                (v s, \, s) & \mbox{if } s  > 0  \\
              \end{array} \right.
              .
  \end{equation}
This is a simple impulse instantaneous change in velocity at $t=0$.

For clock B on the right, we have it remain at rest until the speed of impulse from the left reaches it at the speed of sound.  It then propagates with velocity $v_r$ until it reaches the point where the rod has a length $L+\Delta L$.  Finding this intersection gives the parameter $s_R$.  The resulting parameterization is
\begin{equation}
  (x_B(s),\ t_B(s)) =  \left\{
              \begin{array}{cc}
                 (L, \, s)& \mbox{if } s \leq \frac{L}{v_s}  \\
                (v_r \left(s-\frac{L}{v_s}\right) + L, \, s) & \mbox{if } \frac{L}{v_s} < s  <  s_R \\
                (s v + \sqrt{1-(\frac{v}{c})^2} (L+\Delta L), s) & \mbox{if } s>s_R
              \end{array} \right.
  \end{equation}
where
\begin{equation}
 s_R = \frac{L v_s \left( \frac{\Delta L}{L} \sqrt{1-(\frac{v}{c})^2} + \sqrt{1-(\frac{v}{c})^2} -1\right) + L v_r}{v_s(v_r-v)}.
\end{equation}
At $t=0$ both A and B clocks are synchronized at $0$.
The time on each clock is calculated using
\begin{equation}
\tau = \int_{0}^{s_{\rm{final}}} \, ds \, \sqrt{\left(\frac{dt}{ds}\right)^2 - \frac{1}{c^2} \left(\frac{dx}{ds}\right)^2}.
\end{equation}
Using this the clock A reads
\begin{equation}
 \tau_A = \int_0^{t'/\sqrt{1-(\frac{v}{c})^2}} \, ds \, \sqrt{1-(\frac{v}{c})^2}
\end{equation}
where $t'$ is the time coordinate in the frame of reference with steady velocity $v$.
After the object reaches a steady velocity, clock B reads
\begin{equation}
 \tau_B = \int_0^{L/v_s} ds + \int_{L/v_s}^{s_R}\, ds\,\sqrt{1-(\frac{v_r}{c})^2} + \int_{s_R}^{(t' + \frac{v}{c^2} (L+\Delta L))/\sqrt{1-(\frac{v}{c})^2} }\, ds\,\sqrt{1-(\frac{v}{c})^2}.
\end{equation}
The difference $\tau_B-\tau_A$ is independent of $t'$ and gives eq~\ref{EqDeltaTauGeneral}.  Eq.~\ref{EqTimeDiffSameUniformAcc} follows from a similar approach with the appropriate path parameterizations.

\end{document}